\documentclass[preprint,3p,times,number]{elsarticle}

\usepackage{hyperref}
\usepackage{amsmath}
\usepackage{amssymb}
\usepackage{epsfig}
\usepackage{float}
\usepackage{bm}
\usepackage{epsfig}
\usepackage{soul}
\usepackage{color}
\usepackage{aas_macros}

\newcommand{\ri}{\mathrm{i}}
\newcommand{\re}{\mathrm{e}}
\newcommand{\rd}{\mathrm{d}}

\journal{Physica A}

\begin{document}

\begin{frontmatter}

\title{Functional integral approach to the kinetic theory\\of inhomogeneous systems}

\author[label1]{Jean-Baptiste Fouvry}
\author[label2]{Pierre-Henri Chavanis}
\author[label1,label3]{Christophe Pichon}

\address[label1]{Institut d'Astrophysique de Paris and UPMC, CNRS (UMR 7095), 98 bis Boulevard Arago, 75014, Paris, France}
\address[label2]{Laboratoire de Physique Th\'eorique (IRSAMC), CNRS and UPS, Univ. de Toulouse, F-31062 Toulouse, France}
\address[label3]{Institute of Astronomy \& KICC, University of Cambridge, Madingley Road, Cambridge, CB3 0HA, United Kingdom}

\begin{abstract}

We present a  derivation of the kinetic equation describing the secular
evolution of spatially inhomogeneous systems with long-range interactions,
the so-called inhomogeneous Landau equation, by relying on a functional integral formalism.
We start from the BBGKY hierarchy derived from the Liouville equation. At the
order ${ 1/N }$, where $N$ is the number of particles, the evolution of the
system is characterised by its 1-body distribution function and its 2-body
correlation function. Introducing associated auxiliary fields, the evolution of
these quantities may be rewritten as a traditional functional integral. By
functionally integrating over the 2-body autocorrelation, one obtains a new
constraint connecting the 1-body DF and the auxiliary fields. When inverted,
this constraint allows us to obtain the closed non-linear kinetic equation satisfied by the 1-body distribution function.
This derivation provides an alternative to previous methods, either based on the
direct resolution of the truncated BBGKY hierarchy or on the Klimontovich
equation. It may  turn out to be fruitful to derive more accurate kinetic
equations, e.g. accounting for collective effects, or higher order correlation
terms.

\end{abstract}


\end{frontmatter}

\section{Introduction}
\label{sec:introduction}

Recently, the dynamics and thermodynamics of systems with long-range
interactions has been a subject of active research~\cite{CampaDauxois2009,CampaDauxois2014}. The
equilibrium properties
of these systems, and their specificities such as negative specific heats,
various kinds of phase transitions and ensemble
inequivalence, are now relatively well understood. However, their dynamical
evolution is more complex and many aspects of it need to be improved and
exploited in order to obtain explicit predictions. A short historic of the
early development of kinetic theory for plasmas, stellar systems,
and other systems with long-range
interactions is presented in~\cite{Chavanis2010jsm,Chavanis2012epjp2,Chavanis2013}. The main lines of
this historic are
recalled below, with some complements, in order to place our work in a
general context. We show in particular how the necessity to develop a
kinetic theory for spatially
inhomogeneous systems such as those considered in the present
paper progressively emerged.

The first kinetic theory describing the statistical evolution of a large number
of particles was developed by Boltzmann for a dilute neutral
gas~\cite{Boltzmann1872}. In that
case, the particles do not interact except during strong local collisions.
The gas is spatially homogeneous and the Boltzmann kinetic
equation describes the evolution of the velocity distribution function ${ f(\bm{v},t) }$ of the particles under the effect of strong collisions. It can be shown to satisfy a H-Theorem corresponding to an increase of Boltzmann's entropy.

Boltzmann's kinetic theory was extended to charged gases (plasmas) by
Landau~\cite{Landau1936}.
In that case, the particles interact via long-range Coulombian forces but,
because of electroneutrality and Debye shielding~\cite{DebyeHuckel1923I,DebyeHuckel1923II}, the interaction
is screened
on a lengthscale of the order of the Debye length, so that the collisions are
essentially local. A neutral plasma is spatially homogeneous and the kinetic
equation again describes the evolution of the velocity distribution function
${ f(\bm{v},t) }$ of the charges under the effect of close encounters
(electrostatic deflections). Since these encounters are weak, one can expand the
Boltzmann equation in the limit of small
deflections and make a linear trajectory approximation. This leads to the so-called
Landau equation~\cite{Landau1936} which is valid in such a weak coupling approximation.
The Landau
equation
exhibits a logarithmic divergence at small scales due to the neglect of strong
collisions (that are rare but that cannot be totally neglected) and a
logarithmic divergence at large scales due to the neglect of collective
effects, i.e., the dressing of particles by their polarisation
cloud (because two like sign charges repell each other and two
opposite charges attract each other, a particle of a given charge has the
tendency to be surrounded by a cloud of particles of
opposite charge). Landau regularised these divergences by introducing rather
arbitrarily a lower cut-off at the impact parameter producing a deflection at
$90^{\circ}$
(Landau length) and an upper cut-off at the Debye length. Collective effects
were rigorously taken into account later by
Balescu~\citep{Balescu1960} and Lenard~\citep{Lenard1960}, leading to the
Balescu-Lenard equation. They showed that this equation is valid at the order
${ 1/\Lambda }$, where $\Lambda$ is the plasma parameter (number of charges in the
Debye sphere). The Balescu-Lenard
equation is similar to the
Landau equation except that it includes the square of the dielectric function in
the denominator of the potential of interaction (in Fourier space). The
dielectric function first appeared as a probe of the dynamical
stability of plasmas based on the linearised Vlasov
equation~\cite{Vlasov1938,Vlasov1945}. In the Balescu-Lenard equation the dielectric
function
accounts  for Debye shielding and removes the logarithmic divergence at large
scales present in the Landau equation. This amounts to replacing the bare
potential of interaction by a dressed potential of interaction.
The Landau
equation is recovered
from the Balescu-Lenard equation by replacing the dielectric function by
unity, i.e., by neglecting collective effects. 
In addition to including the
dielectric function, the form of the kinetic equation given by Balescu and
Lenard exhibits a local condition of resonance, encapsulated in a Dirac
${\delta_{\rm D}-}$function. Resonant contributions are the
drivers of diffusion on secular timescales (collisional evolution), as they do
not average out.
When integrating over this condition of resonance, we
recover the original form of the kinetic equation given by Landau.

Self-gravitating systems are spatially inhomogeneous but the early kinetic
theories pioneered by Jeans~\cite{Jeans1929} and Chandrasekhar~\cite{Chandrasekhar1942,Chandrasekhar1943I,Chandrasekhar1943II} were based on the assumption
that the collisions (close encounters) between stars can be treated with a 
local approximation as if the system were infinite and homogeneous. Since a star
experiences a large number of weak deflections, Chandrasekhar
\cite{Chandrasekhar1949} developed an
analogy with Brownian motion. He started from the Fokker-Planck equation and
computed the diffusion and friction coefficients in a binary collision 
theory. This leads to a kinetic equation (usually called the Fokker-Planck
equation by astrophysicists) that is equivalent to the Landau
equation.\footnote{The Landau equation only involves the square of the
potential of interaction, so that it keeps the same form for
Coulombian and gravitational interactions, except for a change in
the prefactor: ${(-e^{2})^{2}}$ has to be replaced by ${ (Gm^{2})^{2} }$.
The kinetic equation derived by Chandrasekhar (see also~\cite{Rosenbluth1957}), albeit physically
equivalent to the Landau equation, did not appear under
the same mathematical form because he started from the Fokker-Planck
equation
${ \partial_{t}
f \!=\! \partial_{v_{i}}\partial_{v_{j}}(D_{ij} f) \!+\! \partial_{v_{i}
}(F^{\rm fric}_{i} \! f ) }$ in which the diffusion tensor is placed after
the two velocity
derivatives, while the Landau
equation can be viewed as a Fokker-Planck equation
${ \partial_{t}
f \!=\! \partial_{v_{i}}(D_{ij}\partial_{v_{j}} f) \!+\! \partial_{v_{i}
}(F^{\rm pol}_{i} \! f ) }$  where the diffusion tensor is
placed between  the two
velocity derivatives. From this second rewriting, Landau obtained a symmetric
expression of the collision operator from which one can directly deduce all the
conservation laws of the system and derive the
$H$-theorem for the Boltzmann entropy. Furthermore, Landau derived simultaneously the
diffusion and friction coefficients, while Chandrasekhar obtained them from two
different calculations and showed a posteriori that they were connected 
at equilibrium by the Einstein relation. Let us
emphasise, however, that the friction force $\bm{F}^{\rm fric}$ computed by 
Chandrasekhar is the true friction force while the friction force $\bm{F}^{\rm
pol}$
appearing in the Landau equation is the friction due to the
polarisation~\cite{Chavanis2013}.}
The gravitational
Landau equation
exhibits a logarithmic divergence at small scales
due to the neglect of strong collisions and a
logarithmic divergence at large scales due to the local approximation
or to the assumption that the system is infinite and homogeneous. Strong
collisions are taken into account in the treatment of Chandrasekhar which
shows, without having to introduce a cut-off,
that the small-scale divergence  is regularised at the gravitational Landau
length. The
large-scale divergence is
usually regularised by introducing a cut-off at the Jeans
length which is the gravitational analogue of the Debye length. The
gravitational Landau equation  is
often thought to be
sufficient to describe the
collisional dynamics of
spherical stellar systems such as globular clusters.
However, the treatment based on the local approximation, or on the assumption
that the system is infinite and homogeneous, is not fully satisfactory since it
leads to a logarithmic divergence. Furthermore, it prevents one from taking
into account collective
effects, i.e., the dressing of stars by polarisation
clouds (because of the  gravitational attraction, a star has
the tendency to be surrounded by a cloud of stars which increases
its effective gravitational mass and reduces its collisional relaxation time).
Indeed, if we naively take
into account
collective effects by introducing the gravitational ``dielectric function'' in
the homogeneous Balescu-Lenard equation (with the sign ${ -G m^{2} }$ instead of
${ +e^{2} }$)
we get a strong, linear, divergence at large scales related to the Jeans
instability of an artificial infinite homogeneous medium. If we
enclose the system within a box, this divergence
suggests that
collective effects accelerate the relaxation (i.e., reduce the relaxation
time) when the size of the
system approaches the Jeans length (see~\cite{Weinberg1993} and  Appendix
E of~\cite{Chavanis2013}). However,
since the size of real stellar systems is precisely of the order of the Jeans
scale where the divergence occurs, this approach is not rigorous and not fully conclusive (no divergence should occur for a stable spatially
inhomogeneous system).

In order to solve these difficulties, the kinetic
theory of stellar systems has recently been generalised to fully inhomogeneous
systems, either when collective
effects are neglected, leading to the inhomogeneous Landau
equation~\citep{Chavanis2010jsm,Chavanis2013}, or when they are taken into
account, leading to
the inhomogeneous Balescu-Lenard equation~\citep{Heyvaerts2010,Chavanis2012}.
These equations are valid at the order ${ 1/N }$, where $N$ is the
number of stars in the system. 
These equations do not present any divergence at large scale since they take
into account the finite extension of the system. They are written with
angle-action variables that are appropriate
to describe the intricate dynamics of stars when the system is spatially
inhomogeneous and multi-periodic.
They also include  a condition of resonance encapuslated in a Dirac
${\delta_{\rm D}-}$function that generalises the one occurring  in the homogeneous 
Balescu-Lenard equation. This condition of resonance is written with
angle-action variables so that it accounts for possibly distant encounters
between stars. This is a crucial difference with plasma physics where the 
encounters between charges are essentially local because of electroneutrality
and Debye shielding. Finally, when collective
effects are taken into account in stellar systems, the inhomogeneous
Balescu-Lenard equation
includes a response matrix written with angle-action variables that generalises
the dielectric function appearing in the  homogeneous Balescu-Lenard
equation of plasma physics. This amounts once again to replacing the bare
potential of interaction by a dressed one.  This effective
potential accounts for anti-shielding (the fact that the gravitational mass of a
star is enhanced by its polarisation cloud) and for the reduction of the
relaxation time. The
inhomogeneous Landau and Balescu-Lenard equations, describing the collisional
evolution of stellar systems, have been recently used in the astrophysical
context~\citep{FouvryPichonChavanis2015,FouvryPichonMagorrianChavanis2015,
FouvryLBThick}, and have proven  fruitful to probe complex secular regimes. In
particular, these works have demonstrated that, in the case of cold stellar
discs, accounting for spatial inhomogeneity and collective effects is crucial
to correctly explain the results of $N$-body simulations~\cite{Sellwood2012}. 
In particular, they clearly established that collective effects cause cool discs
to have shorter two-body relaxation time that one might expect, because each real
particle is accompanied by a cloud of correlated particles.

There are two standard methods to derive kinetic equations for a Hamiltonian
${N-}$body system with long-range interactions. The first one is based on the
Liouville equation for the ${N-}$body distribution function. One writes the first
two equations of the
Bogoliubov-Born-Green-Kirkwood-Yvon (BBGKY) hierarchy and close the hierarchy by
neglecting three-body correlation functions. One then solves the second
equation of the BBGKY hierarchy to express the two-body correlation function in
terms of the one-body distribution function. Subsequently, one substitutes the result in
the first  equation of the BBGKY hierarchy to obtain a self-consistent kinetic
equation. The same results can be obtained by using projection
operator technics. The second method is based on the Klimontovich equation~\citep{Klimontovich1967} for
the discrete distribution function written as a sum of Dirac ${\delta_{\rm D}-}$functions.
One decomposes the exact distribution function into a smooth component plus
fluctuations. One then writes
two evolution equations, one for the smooth component and one for the fluctuations and
closes this system of equations by neglecting nonlinear terms in the equation for
the fluctuations (quasilinear approximation). One then solves the equation for
the
fluctuations and computes the two-body correlation function in terms of the
smooth one-body distribution function. Finally, one substitutes the result
in
the first  equation to obtain  a self-consistent  kinetic equation. These two
methods are physically equivalent, although technically different. It is usually
agreed that the method based on the Klimontovich equation is simpler to implement.

In a little-known seven-page paper,~\cite{JolicoeurGuillou1989} presented a
general functional integral framework suited to the study of classical kinetic
theory. Using this formalism, starting from the Liouville equation, they derived
the entire Bogoliubov-Born-Green-Kirkwood-Yvon (BBGKY) hierarchy. More
interestingly,
they showed in an Appendix how this approach allowed them to derive in a simple
way the homogeneous Balescu-Lenard equation~\citep{Balescu1960,Lenard1960} of
plasma physics. In the present paper, we propose to show how one may use the
functional integral approach introduced in~\citep{JolicoeurGuillou1989} to
derive the inhomogeneous Landau equation, hence presenting a new method to obtain this
kinetic equation. This equation describes the long-term evolution of isolated
stable systems with long-range interactions, which evolve under the effect of
their own discreteness, when collective effects are neglected.  In this
collisional context (i.e., where finite-$N$ effects are taken into account), one
of the main difficulty is to deal with non-local resonances between distant
orbits, as the upcoming calculations will emphasise. Although we have in mind
the application of the kinetic theory to self-gravitating systems (this will
transpire in our presentation), our results are more general, and may find
application for other systems with long-range interactions.

The paper is organised as follows. Section~\ref{sec:BBGKY} sketches a brief derivation of the BBGKY hierarchy.
 Section~\ref{sec:formalism} presents the functional integral formalism from~\cite{JolicoeurGuillou1989}. Section~\ref{sec:application} illustrates how one may use this formalism to derive through a new route the inhomogeneous Landau equation. Finally, section~\ref{sec:conclusion} wraps up.

\section{Derivation of the BBGKY hierarchy}
\label{sec:BBGKY}

In order to introduce the basic equations of the problem, we
first present a brief derivation of the BBGKY hierarchy, following the
notations from~\cite{Chavanis2013}. We
consider a system made of $N$ identical particles, of individual mass ${ \mu
\!=\! M_{\rm tot} /N }$, where $M_{\rm tot}$ is the total mass of the system.
The dynamics of these particles is fully described by Hamilton's equations which
read
\begin{equation}
\mu \frac{\rd \bm{x}_{i}}{\rd t} = \frac{\partial H}{\partial \bm{v}_{i}} \;\;\; ; \;\;\; \mu \frac{\rd \bm{v}_{i}}{\rd t} = - \frac{\partial H}{\partial \bm{x}_{i}} \, ,
\label{Hamiltons_equations}
\end{equation}
where ${ (\bm{x}_{i} , \bm{v}_{i}) }$ stands for the position and velocity
of particle $i$. In equation~\eqref{Hamiltons_equations}, the Hamiltonian $H$ is
given by
\begin{equation}
H = \frac{\mu}{2} \! \sum_{i = 1}^{N} \bm{v}_{i}^{2} + \mu^{2} \! \sum_{i < j} U (|\bm{x}_{i} \!-\! \bm{x}_{j}|) \, ,
\label{Hamiltonian_mass}
\end{equation}
where ${ U (|\bm{x}|)}$ corresponds to the interaction potential, e.g., ${ U (|\bm{x}|) \!=\! - G / |\bm{x}| }$ in the gravitational context. In order to
obtain a statistical description of this system, we may now introduce the ${
N-}$body probability distribution function (PDF) ${ P_{N} (\bm{x}_{1} ,
\bm{v}_{1} , ... , \bm{x}_{N} , \bm{v}_{N} , t) }$ which gives the probability
of finding at time $t$, particle $1$ at position ${ \bm{x}_{1} }$ with
velocity $\bm{v}_{1}$, particle $2$ at position $\bm{x}_{2}$ with
velocity $\bm{v}_{2}$, etc. More precisely, $P_{N}$ is normalised such that
\begin{equation}
\int \!\! \rd \Gamma_{1} \rd \Gamma_{2} ... \rd \Gamma_{N} \, P_{N} (\Gamma_{1} , \Gamma_{2} , ... \Gamma_{N} , t) = 1 \, ,
\label{normalisation_FN}
\end{equation}
where ${ \Gamma_{i} \!=\! (\bm{x}_{i} , \bm{v}_{i} ) }$. The evolution of $P_{N}$ is
governed by Liouville's equation which reads
\begin{equation}
\frac{\partial P_{N}}{\partial t} + \sum_{i = 1}^{N} \bigg[ \bm{v}_{i} \!\cdot\! \frac{\partial P_{N}}{\partial \bm{x}_{i}} + \mu \, \bm{\mathcal{F}}_{\!i}^{\rm tot} \!\cdot\! \frac{\partial P_{N}}{\partial \bm{v}_{i}} \bigg] = 0 \, ,
\label{Liouville_equation}
\end{equation}
where the total force $\bm{\mathcal{F}}_{\!i}^{\rm tot}$ exerted on particle $i$ is given by
\begin{equation}
\bm{\mathcal{F}}_{\!i}^{\rm tot} = \sum_{j \neq i} \bm{\mathcal{F}}_{\!ij} = - \sum_{j \neq i} \frac{\partial U_{ij}}{\partial \bm{x}_{i}} \, .
\label{expression_Fi}
\end{equation}
In equation~\eqref{expression_Fi}, we defined as $\bm{\mathcal{F}}_{\!ij}$ the
force exerted by particle $j$ on particle $i$. Introducing the potential
of interaction ${ U_{ij} \!=\! U (|\bm{x}_{i} \!-\! \bm{x}_{j}|)}$, one has ${
\bm{\mathcal{F}}_{\!ij} \!=\! - \partial U_{ij} / \partial \bm{x}_{i} }$.
It is important to emphasise that the Liouville
equation~\eqref{Liouville_equation} contains the same information as the set of
Hamilton's equations~\eqref{Hamiltons_equations}. One may now define the reduced
PDFs $P_{n}$ for ${ 1 \!\leq\! n \!\leq\! N }$ by
\begin{equation}
P_{n} (\Gamma_{1} , ... , \Gamma_{n} , t) = \!\! \int \!\! \rd \Gamma_{n+1}  ...  \rd \Gamma_{N} \,  P_{N} (\Gamma_{1} , ... , \Gamma_{N} , t) \, .
\label{definition_reduced_DF}
\end{equation}
Using the symmetry of $P_{N}$ w.r.t. permutations of its arguments, one can integrate equation~\eqref{Liouville_equation} w.r.t. to ${ \rd \Gamma_{n+1} ... \rd \Gamma_{N} }$ to obtain the evolution equation satisfied by ${ P_{n} }$. This gives the general term of the BBGKY hierachy which reads
\begin{equation}
\frac{\partial P_{n}}{\partial t} + \sum_{i = 1}^{n} \bm{v}_{i} \!\cdot\! \frac{\partial P_{n}}{\partial \bm{x}_{i}} + \sum_{i = 1}^{n} \sum_{k = 1 , k \neq i}^{n} \, \mu \, \bm{\mathcal{F}}_{\!ik} \!\cdot\! \frac{\partial P_{n}}{\partial \bm{v}_{i}} = - (N \!-\! n) \sum_{i = 1}^{n} \int \rd \Gamma_{n+1} \, \mu \, \bm{\mathcal{F}}_{\!i,n+1} \!\cdot\! \frac{\partial P_{n+1}}{\partial \bm{v}_{i}}  \, .
\label{BBGKY_n}
\end{equation}
Here, one should note that the l.h.s. of equation~\eqref{BBGKY_n} only involves
the first $n$ particles, while the collision term from the r.h.s. involves the
reduced PDF $P_{n+1}$ of higher order, i.e., the hierarchy is not closed. We now
restrict ourselves to the first two  equations of this hierarchy which read
\begin{equation}
\frac{\partial P_{1}}{\partial t} \!+\! \bm{v}_{1} \!\cdot\! \frac{\partial P_{1}}{\partial \bm{x}_{1}} = - (N \!-\! 1) \!\! \int \!\! \rd \Gamma_{2} \, \mu \, \bm{\mathcal{F}}_{\!12} \!\cdot\! \frac{\partial P_{2}}{\partial \bm{v}_{1}} \, ,
\label{BBGKY_1}
\end{equation}
and
\begin{equation}
\frac{\partial P_{2}}{\partial t} \!+\! \bm{v}_{1} \!\cdot\! \frac{\partial P_{2}}{\partial \bm{x}_{1}} \!+\! \bm{v}_{2} \!\cdot\! \frac{\partial P_{2}}{\partial \bm{x}_{2}} \!+\! \mu \, \bm{\mathcal{F}}_{\!12} \!\cdot\! \frac{\partial P_{2}}{\partial \bm{v}_{1}} \!+\! \mu \, \bm{\mathcal{F}}_{\!21} \!\cdot\! \frac{\partial P_{2}}{\partial \bm{v}_{2}} = - (N \!-\! 2) \!\! \int \!\! \rd \Gamma_{3} \, \mu \, \bigg[ \bm{\mathcal{F}}_{\!13} \!\cdot\! \frac{\partial P_{3}}{\partial \bm{v}_{1}} \!+\! \bm{\mathcal{F}}_{\!23} \!\cdot\! \frac{\partial P_{3}}{\partial \bm{v}_{2}} \bigg] \, .
\label{BBGKY_2}
\end{equation}
We may now introduce the reduced distribution functions $f_{n}$ as
\begin{equation}
f_{n} (\Gamma_{1} , ... , \Gamma_{n} , t) = \mu^{n} \frac{N!}{(N \!-\! n)!} P_{n} (\Gamma_{1} , ... , \Gamma_{n} , t) \, .
\label{definition_fn}
\end{equation}
Equations~\eqref{BBGKY_1} and~\eqref{BBGKY_2} immediately take the form
\begin{equation}
\frac{\partial f_{1}}{\partial t} \!+\! \bm{v}_{1} \!\cdot\! \frac{\partial f_{1}}{\partial \bm{x}_{1}} = - \!\! \int \!\! \rd \Gamma_{2} \, \bm{\mathcal{F}}_{\!12} \!\cdot\! \frac{\partial f_{2}}{\partial \bm{v}_{1}} \, ,
\label{BBGKY_1_f}
\end{equation}
and
\begin{equation}
\frac{\partial f_{2}}{\partial t} \!+\! \bm{v}_{1} \!\cdot\! \frac{\partial f_{2}}{\partial \bm{x}_{1}} \!+\! \bm{v}_{2} \!\cdot\! \frac{\partial f_{2}}{\partial \bm{x}_{2}} \!+\! \mu \, \bm{\mathcal{F}}_{\!12} \!\cdot\! \frac{\partial f_{2}}{\partial \bm{v}_{1}} \!+\! \mu \, \bm{\mathcal{F}}_{\!21} \!\cdot\! \frac{\partial f_{2}}{\partial \bm{v}_{2}} = - \!\! \int \!\! \rd \Gamma_{3} \, \bigg[ \bm{\mathcal{F}}_{\!13} \!\cdot\! \frac{\partial f_{3}}{\partial \bm{v}_{1}} \!+\! \bm{\mathcal{F}}_{\!23} \!\cdot\! \frac{\partial f_{3}}{\partial \bm{v}_{2}} \bigg] \, .
\label{BBGKY_2_f}
\end{equation}
In order to emphasise the importance of the correlations between particles, we
define the cluster representation of the reduced distribution functions. We
introduce the  ${2-}$body correlation $g_{2}$ as
\begin{equation}
f_{2} (\Gamma_{1} , \Gamma_{2}) = f_{1} (\Gamma_{1}) f_{1} (\Gamma_{2}) \!+\! g_{2} (\Gamma_{1} , \Gamma_{2}) \, .
\label{definition_g2}
\end{equation}
Similarly, introducing the irreducible ${3-}$body correlation $g_{3}$, one can
express $f_{3}$ as
\begin{equation}
f_{3} (\Gamma_{1} , \Gamma_{2} , \Gamma_{3}) = f_{1} (\Gamma_{1}) f_{1} (\Gamma_{2}) f_{1} (\Gamma_{3}) \!+\! f_{1} (\Gamma_{1}) g_{2} (\Gamma_{2} , \Gamma_{3}) \!+\! f_{1} (\Gamma_{2}) g_{2} (\Gamma_{1} , \Gamma_{3}) \!+\! f_{1} (\Gamma_{3}) g_{2} (\Gamma_{1} , \Gamma_{2}) \!+\! g_{3} (\Gamma_{1} , \Gamma_{2} , \Gamma_{3}) \, .
\label{definition_g3}
\end{equation}
Within this representation, one can study the scalings of the functions $f_{1}$,
$g_{2}$ and $g_{3}$ with the number of particles. Thanks to the definition from
equation~\eqref{definition_reduced_DF}, one has ${ |P_{n}| \!\sim\! 1 }$. Since
${ \mu \!=\! M_{\rm tot} / N }$, the definition from
equation~\eqref{definition_fn} immediately gives ${ |f_{n}| \!\sim\! 1 }$, and in particular
\begin{equation}
\big| f_{1} \big| \!\sim\! 1 \, .
\label{scalings_f1_f2}
\end{equation}
Integrating equation~\eqref{definition_g2} w.r.t. to ${ (\Gamma_{1} ,
\Gamma_{2}) }$, we obtain ${ \int\!\! \rd \Gamma_{1} \rd
\Gamma_{2} \, g_{2} (\Gamma_{1} , \Gamma_{2}) \!=\! - \mu^{2} N  }$. Similarly,
integrating equation~\eqref{definition_g3} w.r.t. ${ (\Gamma_{1} , \Gamma_{2} ,
\Gamma_{3}) }$, we obtain ${ \int \!\! \rd \Gamma_{1} \rd
\Gamma_{2} \rd \Gamma_{3} \, g_{3} (\Gamma_{1} , \Gamma_{2} ,
\Gamma_{3})  \!=\! 2 \mu^{3} N }$. As a
consequence, we have the scalings
\begin{equation}
\big| g_{2} \big| \!\sim\!  \frac{1}{N} \;\;\; ; \;\;\; \big| g_{3} \big|  \!\sim\! \frac{1}{N^{2}} \, .
\label{scalings_g2_g3}
\end{equation}
Using the decompositions from equations~\eqref{definition_g2} and~\eqref{definition_g3}, after some simple calculations, one can rewrite equations~\eqref{BBGKY_1_f} and~\eqref{BBGKY_2_f} as
\begin{equation}
\frac{\partial f_{1}}{\partial t} \!+\! \bm{v}_{1} \!\cdot\! \frac{\partial f_{1}}{\partial \bm{x}_{1}} \!+\! \bigg[ \! \int \!\! \rd \Gamma_{2} \, \bm{\mathcal{F}}_{\!12} f_{1} (\Gamma_{2}) \bigg] \!\cdot\! \frac{\partial f_{1}}{\partial \bm{v}_{1}} = -  \frac{\partial }{\partial \bm{v}_{1}} \!\cdot\! \bigg[ \!\! \int \!\! \rd \Gamma_{2} \,  \bm{\mathcal{F}}_{\!12} \, g_{2} (\Gamma_{1} , \Gamma_{2} )  \bigg] \, ,
\label{BBGKY_1_f_rewrite}
\end{equation}
and
\begin{align}
& \, \frac{1}{2} \frac{\partial g_{2}}{\partial t} \!+\! \bm{v}_{1} \!\cdot\! \frac{\partial g_{2}}{\partial \bm{x}_{1}} + \, \bigg[ \! \int \!\! \rd \Gamma_{3} \, \bm{\mathcal{F}}_{\!13} f_{1} (\Gamma_{3}) \bigg] \!\cdot\! \frac{\partial g_{2}}{\partial \bm{v}_{1}} + \mu \, \bm{\mathcal{F}}_{\!12} \!\cdot\! \frac{\partial f_{1}}{\partial \bm{v}_{1}} f_{1} (\Gamma_{2}) + \bigg[ \! \int \!\! \rd \Gamma_{3} \, \bm{\mathcal{F}}_{\!13} \, g_{2} (\Gamma_{2} , \Gamma_{3})  \bigg] \!\cdot\! \frac{\partial f_{1}}{\partial \bm{v}_{1}} \nonumber
\\
& \, \!+\! \mu \, \bm{\mathcal{F}}_{\!12} \!\cdot\! \frac{\partial g_{2}}{\partial \bm{v}_{1}} \!+\! \frac{\partial }{\partial \bm{v}_{1}} \!\cdot\! \bigg[ \!\! \int \!\! \rd \Gamma_{3} \bm{\mathcal{F}}_{\!13} \, g_{3} (\Gamma_{1} , \Gamma_{2} , \Gamma_{3}) \bigg] \, + (1 \leftrightarrow 2) = 0 \, .
\label{BBGKY_2_f_rewrite}
\end{align}
We may now perform a truncation at the order ${1/N}$ of the two
equations~\eqref{BBGKY_1_f_rewrite} and~\eqref{BBGKY_2_f_rewrite}. To do so, we
rely on the scalings from equations~\eqref{scalings_f1_f2}
and~\eqref{scalings_g2_g3}, and on the fact that ${ \mu \!\sim\! 1/N }$ and ${ |
\bm{\mathcal{F}}_{\!12} | \!\sim\! 1 }$. In equation~\eqref{BBGKY_1_f_rewrite},
all the terms are of order ${1/N}$ or larger so that they should all be
kept. In equation~\eqref{BBGKY_2_f_rewrite}, one can note that the terms from
the first line are all of order ${1/N}$ and have to be conserved, while all
the terms from the second line are of order ${ 1/N^{2} }$, and may therefore
be neglected.\footnote{There is, however, a subtlety with the first term on the
second line of equation~\eqref{BBGKY_2_f_rewrite}. Indeed, even if this term is
of order ${ 1/N^{2} }$ ``in average'', it can become very large when particle
$2$ approaches particle $1$ due to the divergence of the Coulombian or
gravitational force at small separations. This term accounts for the
effect of strong collisions. Even if strong collisions are not
dominant for systems with long-range interactions, they have to be taken into
account for 3D Coulombian or gravitational systems otherwise a logarithmic
divergence occurs at small scales. This implies that the ${ 1/N }$ expansion is not
uniformly convergent. More details can be found in~\cite{Chavanis2013}. In
the present paper, for the sake of simplicity, we shall ignore this difficulty.}
In addition to these truncations, and in order to consider
quantities of order $1$, we finally introduce the ${1-}$body DF $F$ and the
${2-}$body correlation function $\mathcal{C}$ as
\begin{equation}
F = f_{1} \;\;\; ; \;\;\; \mathcal{C} = \frac{g_{2}}{\mu} \, .
\label{definition_F1_C2}
\end{equation}
It is straighforward to note that these functions scale like ${ | F |
\!\sim\! 1 }$ and ${ | \mathcal{C} | \!\sim\! 1 }$. Using these new
functions, the first two equations~\eqref{BBGKY_1_f_rewrite}
and~\eqref{BBGKY_2_f_rewrite} of the BBGKY hierarchy when truncated at the order
${1/N}$  take the form
\begin{equation}
\frac{\partial F}{\partial t} \!+\! \bm{v}_{1} \!\cdot\! \frac{\partial
F}{\partial \bm{x}_{1}} + \!\bigg[ \! \int \!\! \rd \Gamma_{2} \,
\bm{\mathcal{F}}_{\!12} F (\Gamma_{2}) \bigg] \!\cdot\! \frac{\partial
F}{\partial \bm{v}_{1}} = - \, \mu \,  \frac{\partial }{\partial \bm{v}_{1}}
\!\cdot\! \bigg[ \! \int \!\! \rd \Gamma_{2} \, \bm{\mathcal{F}}_{\!12}
\, \mathcal{C} (\Gamma_{1} , \Gamma_{2}) \bigg] \, ,
\label{BBGKY_1_final}
\end{equation}
and
\begin{align}
& \, \frac{1}{2} \frac{\partial \mathcal{C}}{\partial t} \!+\! \bm{v}_{1} \!\cdot\! \frac{\partial \mathcal{C}}{\partial \bm{x}_{1}} \!+\! \bigg[ \! \int \!\! \rd \Gamma_{3} \, \bm{\mathcal{F}}_{\!13} F (\Gamma_{3}) \bigg] \!\cdot\! \frac{\partial \mathcal{C}}{\partial \bm{v}_{1}} + \bm{\mathcal{F}}_{\!12} \!\cdot\! \frac{\partial F}{\partial \bm{v}_{1}} F (\Gamma_{2}) \!+\! \bigg[ \! \int \!\! \rd \Gamma_{3} \, \bm{\mathcal{F}}_{\!13} \, \mathcal{C} (\Gamma_{2} , \Gamma_{3}) \bigg] \!\cdot\! \frac{\partial F}{\partial \bm{v}_{1}} + (1 \leftrightarrow 2) = 0 \, .
\label{BBGKY_2_final}
\end{align}
These two evolution equations, which only involve $F$ and $\mathcal{C}$, are the
two coupled equations which are central to the upcoming  functional integral
formalism. The physical interpretation of the terms appearing in these
equations can be found in~\cite{Chavanis2013}.

\section{Functional integral formalism}
\label{sec:formalism}

\cite{JolicoeurGuillou1989} relied on the general functional integral
formalism~\citep{FaddeevSlavnov1993} to derive the entire BBGKY hierarchy as
well as the homogeneous Balescu-Lenard equation for plasma physics. The main
result used is the following one. Let us consider a dynamical
quantity $f$ depending on the time $t$ and defined on a generic phase-space
${ \Gamma \!=\! (\bm{q} , \bm{p}) }$. 
We assume that this quantity evolves according to an equation of the
form
\begin{equation}
[ \partial_{t} \!+\! L ] f \!=\! 0 \, ,
\label{definition_L}
\end{equation}
where $L$ is a differential operator. We denote as $f_{0}$ the solution of equation~\eqref{definition_L}. Our starting point is to rewrite the dynamical constraint from equation~\eqref{definition_L} under a functional integral of the form
\begin{equation}
1 = \!\! \int \!\! \mathcal{D} f \, \delta_{D} (f \!-\! f_{0}) \, .
\label{rewrite_functional}
\end{equation}
We recall that the composition of a function and a ${\delta_{\rm D}-}$functional satisfies
\begin{align}
\delta_{\rm D} ([\partial_{t} \!+\! L] f) & \, = \delta_{\rm D} ([\partial_{t}
\!+\! L] (f \!-\! f_{0})) \nonumber
\\
& \, = \frac{\delta_{\rm D} (f \!-\! f_{0})}{\det \big| \partial ([\partial_{t} \!+\! L] f) / \partial f \big|} \, .
\label{composition_delta}
\end{align}
As the determinant appearing in equation~\eqref{composition_delta} is only a pure number, independent of the dynamical quantity $f$, it may be dropped in the next calculations. As a consequence, equation~\eqref{rewrite_functional} becomes
\begin{equation}
1 = \!\! \int \!\! \mathcal{D} f \, \delta_{\rm D} ([\partial_{t} \!+\! L] f) \, . 
\label{rewrite_functional_II}
\end{equation}
Finally, we recall that the ${\delta_{\rm D}-}$functional satisfies the general
identity
\begin{equation}
\delta_{\rm D} (g) = \!\! \int \!\! \mathcal{D} \lambda \exp \bigg[ \ri \!\! \int \!\! \rd t \rd \Gamma \, \lambda \, g \bigg] \, ,
\label{identity_delta}
\end{equation}
where the auxiliary field $\lambda$ is defined on the same space as $g$. As a consequence, the evolution constraint on $f$ from equation~\eqref{rewrite_functional_II} may be rewritten under the form
\begin{equation}
1 = \!\! \int \!\! \mathcal{D} f \, \mathcal{D} \lambda \, \exp \bigg\{
\ri \!\! \int \!\! \rd t \, \rd \Gamma \, \lambda \, \big[
\partial_{t} \!+\! L \big] f \bigg\} \, .
\label{functional_integration}
\end{equation}
In analogy with the classical limit of quantum field theory, the argument of the
exponential ${ S \![f, \lambda ] \!=\! \ri \!\! \int \!\! \rd t \rd \Gamma \lambda [\partial_{t} \!+\! L] f }$ in equation~\eqref{functional_integration} is called the
action\footnote{It should not be mixed up with the action coordinates from
inhomogeneous dynamics, see section~\ref{sec:angleaction}.},
while equation~\eqref{functional_integration} is the corresponding classical path integral. Finally, in equation~\eqref{functional_integration}, one can note that the evolution equation of $f$ corresponds to the quantity by which the auxiliary field $\lambda$ is multiplied in the action.

In the present paper, we are interested in the long-term collisional evolution of an inhomogeneous system made of $N$ particles. As detailed in section~\ref{sec:BBGKY}, to describe such a system, one has to consider simultaneously two dynamical quantities, namely the ${1-}$body distribution function (DF) ${ F (t , \Gamma) }$ and the ${ 2-}$body autocorrelation function ${ \mathcal{C}  (t , \Gamma_{1} , \Gamma_{2}) }$. Here $F$ satisfies the normalisation constraint
\begin{equation}
\int \!\! \rd \Gamma \, F (t , \Gamma) = N \, \mu = M_{\rm tot} \, ,
\label{normalisation_DF}
\end{equation}
where $M_{\rm tot}$ is the total mass of the system, and ${ \mu \!=\! M_{\rm tot} / N }$
is the mass of the individual particles. As presented in
section~\ref{sec:BBGKY}, at first-order in ${ \varepsilon \!=\! 1/N }$, the
evolution of the system is entirely characterised by the dynamical quantities
$F$
and $\mathcal{C}$. The truncated first two equations of the BBGKY
hierarchy~\eqref{BBGKY_1_final} and~\eqref{BBGKY_2_final} then form a pair of
coupled evolution equations which describe the simultaneous evolutions of these
dynamical quantities. Introducing the auxiliary fields ${ \lambda_{1} (t ,
\Gamma_{1}) }$ and ${ \lambda_{2} (t , \Gamma_{1} , \Gamma_{2}) }$ respectively
associated with $F$ and $\mathcal{C}$, these coupled evolution equations may
straightforwardly  be rewritten under the functional form
\begin{equation}
1 = \!\! \int \!\! \mathcal{D} F \, \mathcal{D} \mathcal{C} \, \mathcal{D} \lambda_{1} \, \mathcal{D} \lambda_{2} \, \exp \bigg\{ \ri \, \bigg[ \! \int \!\! \rd t \, \rd \Gamma_{1} \, \lambda_{1} (A_{1} F \!+\! B_{1} \mathcal{C}) \!+\! \frac{1}{2} \! \int \!\! \rd t \, \rd \Gamma_{1} \, \rd \Gamma_{2} \, \lambda_{2} (A_{2} \mathcal{C} \!+\! D_{2} \mathcal{C} \!+\! S_{2}) \bigg] \bigg\} \, .
\label{functional_all_initial}
\end{equation}
In equation~\eqref{functional_all_initial}, the operators $A_{1}$, $B_{1}$,
$A_{2}$, $D_{2}$ and $S_{2}$ (see equations~\eqref{BBGKY_1_final}
and~\eqref{BBGKY_2_final}) are given by
\begin{align}
A_{1} F & \,  = \bigg[ \frac{\partial }{\partial t} \!+\! \bm{v}_{1} \!\cdot\! \frac{\partial }{\partial \bm{x}_{1}} \!+\! \bigg[ \!\! \int \!\! \rd \Gamma_{2} \, \bm{\mathcal{F}}_{\!12} \, F (\Gamma_{2}) \bigg]  \!\cdot\! \frac{\partial }{\partial \bm{v}_{1}} \bigg] \, F (\Gamma_{1}) \, , \nonumber
\\
B_{1} \mathcal{C} & \, = \mu \!\! \int \!\! \rd \Gamma_{2} \, \bm{\mathcal{F}}_{\!12} \!\cdot\! \frac{\partial \mathcal{C} (\Gamma_{1} , \Gamma_{2}) }{\partial \bm{v}_{1}} \, , \nonumber
\\
A_{2} \mathcal{C} & \, = \bigg[ \frac{\partial }{\partial t} \!+\! \bm{v}_{1} \!\cdot\! \frac{\partial }{\partial \bm{x}_{1}} \!+\! \bm{v}_{2} \!\cdot\! \frac{\partial }{\partial \bm{x}_{2}} \!+\!\! \int \!\! \rd \Gamma_{3} \, F (\Gamma_{3}) \, \bigg[ \bm{\mathcal{F}}_{\!13} \!\cdot\! \frac{\partial }{\partial \bm{v}_{1}} \!+\! \bm{\mathcal{F}}_{\!23} \!\cdot\! \frac{\partial }{\partial \bm{v}_{2}} \bigg] \bigg] \, \mathcal{C} (\Gamma_{1}, \Gamma_{2}) \, ,\nonumber
\\
D_{2} \mathcal{C} & \, = \bigg[ \!\! \int \!\! \rd \Gamma_{3} \, \bm{\mathcal{F}}_{\!13} \, \mathcal{C} (\Gamma_{2} , \Gamma_{3}) \bigg] \!\cdot\! \frac{\partial F}{\partial \bm{v}_{1}} \!+\! (1 \!\leftrightarrow 2) \, , \nonumber
\\
S_{2} & \, =  F (\Gamma_{2}) \, \bm{\mathcal{F}}_{\!12} \!\cdot\! \frac{\partial
F}{\partial \bm{v}_{1}} + (1 \!\leftrightarrow\! 2 ) \, ,
\label{definition_A_B}
\end{align}
where we represented our phase space canonical variables $\Gamma$ as ${ \Gamma
\!=\! (\bm{x} , \bm{v}) }$, and did not write explicitly the dependence w.r.t. $t$
so as to simplify the notations. In the expression of ${ B_{1} \mathcal{C}
}$, one should note the presence of the small factor ${ \mu \!=\! M_{\rm tot} /
N }$, illustrating the fact that we are considering a kinetic development at
first order in ${ \varepsilon \!=\! 1/N }$.
Finally, in equation~\eqref{functional_all_initial}, one may note the presence
of a factor ${1/2}$ in front of the second action term. This was only added for
later convenience; it does not play any role on the final expression of the
evolution equations since it was added as a global prefactor to the constraints
given by the dynamical equations. One can now detail the physical content of
each of the terms appearing in equation~\eqref{functional_all_initial}. Here,
$A_{1} F$ corresponds to the usual ${1-}$body Vlasov advection term, $B_{1}
\mathcal{C}$ is a first-order term (because of the presence of the mass factor
${ \mu \!=\! M_{\rm tot} / N }$) which corresponds to the ${ 1/N-}$sourcing of
the ${ 1-}$body DF's evolution under the effect of the ${2-}$body autocorrelation
$\mathcal{C}$. Similarly, $A_{2} \mathcal{C}$ corresponds to the usual
${2-}$body Vlasov advection term, $D_{2} \mathcal{C}$ corresponds to the
collective effects, e.g., the Debye shielding for plasmas or  the self-gravity
for stellar systems, while $S_{2}$ is a source term, depending only on $F$,
which sources the dynamics of the ${2-}$body autocorrelation.

In order to obtain a closed kinetic equation describing the long term evolution
of $F$, the traditional approach~\citep{Balescu1960,Lenard1960}, as discussed
above, is the following one. One can first integrate
equation~\eqref{functional_all_initial} functionally over $\lambda_{2}$. As in
equation~\eqref{functional_integration}, this gives an evolution constraint ${
(A_{2} \mathcal{C} \!+\! D_{2} \mathcal{C} \!+\! S_{2}) \!=\!0 }$ which couples
$\mathcal{C}$ and $F$. One must then invert this equation so as to obtain ${
\mathcal{C} \!=\! \mathcal{C} [F] }$. By functionally integrating
equation~\eqref{functional_all_initial} w.r.t. to $\lambda_{1}$, one obtains an
additional evolution constraint ${ ( A_{1} \!+\! B_{1} \mathcal{C})  \!=\! 0 }$,
which involves both $F$ and $\mathcal{C}$. Injecting the previously obtained
expression of $\mathcal{C}$ in this new constraint, one finally obtains a closed
evolution equation involving $F$ only: this is the Balescu-Lenard equation.

However, thanks to the functional rewriting from
equation~\eqref{functional_all_initial},~\cite{JolicoeurGuillou1989} suggested a
different  strategy. This is based on a rewriting of
equation~\eqref{functional_all_initial} under the form
\begin{align}
1 \!=\! \!\! \int \!\!\! \mathcal{D} F \mathcal{D} \mathcal{C}\mathcal{D}
\lambda_{1} \mathcal{D} \lambda_{2}  \exp \bigg\{ & \, \ri \!\! \int
\!\!\! \rd t \, \rd \Gamma_{1} \, \lambda_{1}(\Gamma_{1}) \, A_{1}
F(\Gamma_{1}) \!+\! \frac{\ri}{2} \!\! \int \!\!\!
\rd t \, \rd \Gamma_{1} \rd \Gamma_{2} \,
\lambda_{2}(\Gamma_{1} , \Gamma_{2}) \, G (\Gamma_{1} , \Gamma_{2})  \nonumber
\\
& \, \!-\! \frac{\ri}{2} \!\! \int \!\!\! \rd t \,\rd \Gamma_{1} \rd \Gamma_{2} \, \mathcal{C}(\Gamma_{1} , \Gamma_{2}) \, E (\Gamma_{1} , \Gamma_{2}) \bigg\} \, ,
\label{rewriting_functional_all}
\end{align}
where we defined the quantity ${ G(\Gamma_{1} , \Gamma_{2}) }$ as
\begin{equation}
G (\Gamma_{1} , \Gamma_{2}) = \bm{\mathcal{F}}_{\!12} \!\cdot\! \bigg[ F(\Gamma_{2}) \frac{\partial F}{\partial \bm{v}_{1}} \!-\! F (\Gamma_{1}) \frac{\partial F}{\partial \bm{v}_{2}} \bigg] \, ,
\label{definition_G}
\end{equation}
which corresponds to the contribution from the source term ${ \lambda_{2} S_{2} }$ in equation~\eqref{functional_all_initial}, from which ${ \mathcal{C} }$ is absent.
In equation~\eqref{rewriting_functional_all}, we also introduced ${ E (\Gamma_{1} , \Gamma_{2}) }$ as
\begin{equation}
E (\Gamma_{1} , \Gamma_{2}) = \, A_{2} \, \lambda_{2} (\Gamma_{1} , \Gamma_{2}) \!+\! \!\! \int \!\! \rd \Gamma_{3} \, \bigg[ \bm{\mathcal{F}}_{\!13} \lambda_{2} (\Gamma_{2} , \Gamma_{3} ) \!+\! \bm{\mathcal{F}}_{\!23} \lambda_{2} (\Gamma_{1} , \Gamma_{3}) \bigg] \!\cdot\! \frac{\partial F}{\partial \bm{v}_{3}} + \mu \, \bm{\mathcal{F}}_{\!12} \!\cdot\! \bigg[ \frac{\partial \lambda_{1}}{\partial \bm{v}_{1}} \!-\! \frac{\partial \lambda_{1}}{\partial \bm{v}_{2}} \bigg] \, .
\label{definition_E}
\end{equation}
The three terms present in equation~\eqref{definition_E} can straightforwardly
be obtained from equation~\eqref{functional_all_initial} through the following
manipulations. The first term comes from the component ${ \lambda_{2} A_{2}
\mathcal{C} }$ in equation~\eqref{functional_all_initial}. One has to use an
integration by parts and get rid of the boundary terms. To invert the time
derivative with ${ t \!\in\! [ 0 ; T] }$, where $T$ is an arbitrary temporal
upper bound, we assume that ${ \mathcal{C} (t \!=\! 0) \!=\! 0 }$ (the system is
supposed to be initially uncorrelated) and ${ \lambda_{2} (T) \!=\! 0 }$ (we are
free to impose a condition on $\lambda_{2}$). The second term comes from the
component ${ \lambda_{2} D_{2} \mathcal{C} }$ in
equation~\eqref{functional_all_initial}, where the only operation required is to
permute accordingly the indices ${(1,2,3)}$. Finally, the third term comes from
${ \lambda_{1} {B}_{1} \mathcal{C} }$ in
equation~\eqref{functional_all_initial}. One has to use the integration by parts
formula, get rid of the boundary terms, and use the fact that ${
\bm{\mathcal{F}}_{\!12} }$ is independent of $\bm{v}_{1}$ so that ${ \partial /
\partial \bm{v}_{1} \!\cdot\! [ \lambda_{1} \bm{\mathcal{F}}_{\!12} ] \!=\!
\bm{\mathcal{F}}_{12} \!\cdot\! \partial \lambda_{1} / \partial \bm{v}_{1} }$.
One also has to use the permutation ${ 1 \!\leftrightarrow\! 2 }$, for which ${
\bm{\mathcal{F}}_{\!12} \!=\! - \bm{\mathcal{F}}_{\!21} }$, and recovers the
factor ${ 1/2 }$ present in equation~\eqref{rewriting_functional_all}. At this
stage, it is crucial to note that in the rewriting of
equation~\eqref{rewriting_functional_all} all the dependences on $\mathcal{C}$
have been put in the prefactor ${ \mathcal{C} (\Gamma_{1} , \Gamma_{2}) }$
multitplying the quantity ${ E(\Gamma_{1} , \Gamma_{2}) }$.

\cite{JolicoeurGuillou1989} then suggested the following steps. By first integrating functionally equation~\eqref{rewriting_functional_all} w.r.t. to $\mathcal{C}$, one obtains a new dynamical constraint ${ E [F , \lambda_{1} , \lambda_{2}] \!=\! 0  }$. This constraint may then be inverted so as to obtain ${ \lambda_{2} \!=\! \lambda_{2} [F , \lambda_{1}] }$. The final step of the calculation is then to make this substitution in equation~\eqref{rewriting_functional_all}, which now only involves $\lambda_{1}$ and $F$. By functionally integrating it w.r.t. $\lambda_{1}$, one obtains a closed kinetic equation involving $F$ only: this is the Balescu-Lenard equation.

In their appendix,~\cite{JolicoeurGuillou1989} explicitly applied this
new strategy to derive the homogeneous Balescu-Lenard equation, and showed that
this approach was not only succesful but fairly simple. In the present paper, we
intend to show how one may use the same strategy in the inhomogeneous context.
In order to simplify the calculations, we will neglect collective effects and
show how one can then recover the inhomogeneous Landau equation (the reduced
form of the inhomogeneous Balescu-Lenard equation when collective effects are
neglected). The generalisation of these calculations to the case where
collective effects are take into account will be the subject of a future work.

\section{Application to inhomogeneous systems}
\label{sec:application}

\subsection{Angle-action coordinates}
\label{sec:angleaction}

When considering an inhomogeneous system, the trajectories of the particles
tend to be fairly intricate. We therefore restrict ourselves to symmetric
configurations for which the mean gravitational background potential $\psi_{0}$
associated with the Hamiltonian $H_{0}$ is quasi-stationary and integrable. As
a consequence, one can always remap the physical phase-space coordinates ${
(\bm{x} , \bm{v}) }$ to the angle-action ones ${ (\bm{\theta} , \bm{J})
}$~\citep{Goldstein1950,Born1960,BinneyTremaine2008}. The intrinsic frequencies
of motion along the action torus are defined as
\begin{equation}
\bm{\Omega} (\bm{J}) = \dot{\bm{\theta}} = \frac{\partial H_{0}}{\partial \bm{J}} \, .
\label{definition_Omega}
\end{equation}
Within these new coordinates, along the unperturbed orbits, the angles
$\bm{\theta}$ are ${ 2 \pi-}$periodic, evolving with the frequencies
$\bm{\Omega}$, while the actions $\bm{J}$ are conserved. This change of
coordinates is canonical so that the infinitesimal volumes are conserved, i.e.,
\begin{equation}
\rd \Gamma = \rd \bm{x} \rd \bm{v} = \rd \bm{\theta} \rd \bm{J} \, .
\label{canonical_transformation}
\end{equation}
Relying on the adiabatic
approximation~\citep{Heyvaerts2010,Chavanis2012,Chavanis2013}, we assume that
the ${ 1-}$body DF $F$ evolves in a quasi-stationary fashion,
so that ${ F (\bm{\theta} , \bm{J}) \!=\! F (\bm{J}) }$, where the
dependence w.r.t. $t$ has not been written out explicitly to simplify the notations.
Since $\lambda_{1}$ is the auxiliary field associated with $F$, we also have ${
\lambda_{1} (\bm{\theta} , \bm{J}) \!=\! \lambda_{1} (\bm{J}) }$. The second
auxiliary field ${ \lambda_{2} (\bm{\theta}_{1} , \bm{J}_{1} , \bm{\theta}_{2} ,
\bm{J}_{2}) }$ remains fully dependent on the angle-action coordinates,
contrary to the assumption ${ \lambda_{2} ( \bm{x}_{1} \!-\! \bm{x}_{2}
, \bm{v}_{1} , \bm{v}_{2} ) }$ used in the homogeneous
case~\cite{JolicoeurGuillou1989}. Another property of these coordinates comes
from the derivatives along the mean motion which take the simple form
\begin{equation}
\bm{v}_{1} \!\cdot\! \frac{\partial }{\partial \bm{x}_{1}} + \bigg[ \!\! \int \!\! \rd \Gamma_{2} \bm{\mathcal{F}}_{\!12} F (\Gamma_{2}) \bigg] \!\cdot\! \frac{\partial }{\partial \bm{v}_{1}} = \bm{\Omega}_{1} \!\cdot\! \frac{\partial }{\partial \bm{\theta}_{1}} \, .
\label{derivative_motion}
\end{equation}
Finally, we will also rely on the invariance of the Poisson bracket under the
change of coordinates ${ (\bm{x} , \bm{v}) \!\mapsto\! (\bm{\theta} , \bm{J})
}$, so that for any functions ${ L_{1} (\bm{x} , \bm{v}) }$ and ${ L_{2} (\bm{x} ,
\bm{v}) }$, one may write
\begin{equation}
\frac{\partial L_{1}}{\partial \bm{x}} \!\cdot\! \frac{\partial L_{2}}{\partial
\bm{v}}-\frac{\partial L_{1}}{\partial \bm{v}} \!\cdot\! \frac{\partial
L_{2}}{\partial \bm{x}} = \frac{\partial L_{1}}{\partial \bm{\theta}} \!\cdot\!
\frac{\partial L_{2}}{\partial \bm{J}} \!-\! \frac{\partial L_{1}}{\partial
\bm{J}} \!\cdot\! \frac{\partial L_{2}}{\partial \bm{\theta}} \, .
\label{invariance_Poisson_bracket}
\end{equation}
Using these transformations, one can rewrite in angle-action space the quantities appearing in equation~\eqref{rewriting_functional_all}. Since we assumed that the ${1-}$body DF is quasi-stationary, one has ${ \partial F / \partial \bm{\theta} \!=\! 0 }$, so that thanks to equation~\eqref{derivative_motion}, equation~\eqref{definition_A_B} gives
\begin{equation}
A_{1} F = \frac{\partial F}{\partial t} \, .
\label{A1_action}
\end{equation}
Similarly, thanks to equation~\eqref{invariance_Poisson_bracket},  the quantity $G$ from equation~\eqref{definition_G} may be rewritten as
\begin{equation}
G (\Gamma_{1} , \Gamma_{2}) = - \, \bigg[ F(\bm{J}_{2}) \frac{\partial U_{12}}{\partial \bm{\theta}_{1}} \!\cdot\! \frac{\partial F}{\partial \bm{J}_{1}} + F (\bm{J}_{1}) \frac{\partial U_{21}}{\partial \bm{\theta}_{2}} \!\cdot\! \frac{\partial F}{\partial \bm{J}_{2}} \bigg] \, .
\label{G_action}
\end{equation}
Finally, using the fact that the auxiliary field $\lambda_{1}$ is a quasi-stationary quantity such that ${ \lambda_{1} (\Gamma) \!=\! \lambda_{1} (\bm{J}) }$, we may rewrite the constraint ${ E (\Gamma_{1} , \Gamma_{2}) }$ from equation~\eqref{definition_E} as
\begin{align}
E (\Gamma_{1} , \Gamma_{2}) = & \, \frac{\partial \lambda_{2}}{\partial t} \!+\! \bm{\Omega}_{1} \!\cdot\! \frac{\partial \lambda_{2}}{\partial \bm{\theta}_{1}} \!+\! \bm{\Omega}_{2} \!\cdot\! \frac{\partial \lambda_{2}}{\partial \bm{\theta}_{2}}  \nonumber
\\
& \, +  \!\! \int \!\! \rd \Gamma_{3} \, \bigg[ \frac{\partial U_{31}}{\partial \bm{\theta}_{3}} \!\cdot\! \frac{\partial F}{\partial \bm{J}_{3}} \lambda_{2} (\Gamma_{2} , \Gamma_{3}) \!+\! \frac{\partial U_{32}}{\partial \bm{\theta}_{3}} \!\cdot\! \frac{\partial F}{\partial \bm{J}_{3}} \lambda_{2} (\Gamma_{1} , \Gamma_{3}) \bigg] \nonumber
\\
& \, - \, \mu \, \bigg[ \frac{\partial U_{12}}{\partial \bm{\theta}_{1}} \!\cdot\! \frac{\partial \lambda_{1}}{\partial \bm{J}_{1}} \!+\! \frac{\partial U_{21}}{\partial \bm{\theta}_{2}} \!\cdot\! \frac{\partial \lambda_{1}}{\partial \bm{J}_{2}} \bigg] \, .
\label{E_action}
\end{align}
As presented in~\citep{Chavanis2013}, we will now neglect collective effects,
i.e., neglect contributions associated with the term ${ D_{2} \mathcal{C} }$ in
equation~\eqref{definition_A_B}. Under these
conditions, equation~\eqref{E_action} becomes
\begin{equation}
E (\Gamma_{1} , \Gamma_{2}) = \frac{\partial \lambda_{2}}{\partial
t} + \bm{\Omega}_{1} \!\cdot\! \frac{\partial \lambda_{2}}{\partial
\bm{\theta}_{1}} + \bm{\Omega}_{2} \!\cdot\! \frac{\partial
\lambda_{2}}{\partial \bm{\theta}_{2}} - \, \mu \, \bigg[ \frac{\partial U_{12}}{\partial
\bm{\theta}_{1}} \!\cdot\! \frac{\partial \lambda_{1}}{\partial \bm{J}_{1}} +
\frac{\partial U_{21}}{\partial \bm{\theta}_{2}} \!\cdot\! \frac{\partial
\lambda_{1}}{\partial \bm{J}_{2}} \bigg] \, ,
\label{E_action_nodressing}
\end{equation}
and the constraint ${ E (\Gamma_{1} , \Gamma_{2}) \!=\! 0 }$ implies
\begin{equation}
\frac{\partial \lambda_{2}}{\partial t} + \bm{\Omega}_{1} \!\cdot\! \frac{\partial \lambda_{2}}{\partial \bm{\theta}_{1}} + \bm{\Omega}_{2} \!\cdot\! \frac{\partial \lambda_{2}}{\partial \bm{\theta}_{2}} \!-\! \, \mu \, \bigg[ \frac{\partial U_{12}}{\partial \bm{\theta}_{1}} \!\cdot\! \frac{\partial \lambda_{1}}{\partial \bm{J}_{1}} + \frac{\partial U_{21}}{\partial \bm{\theta}_{2}} \!\cdot\! \frac{\partial \lambda_{1}}{\partial \bm{J}_{2}} \bigg] = 0 \, .
\label{E_0_rewriting}
\end{equation}

\subsection{Inverting the constraint}
\label{sec:inversionconstraint}

In order to solve equation~\eqref{E_0_rewriting}, we will rely
on Bogoliubov's Ansatz (adiabatic approximation). We assume that the fluctuations (i.e.,
$\lambda_{2}$) evolve rapidly compared to the mean dynamical quantities (i.e.,
$F$ and $\lambda_{1}$). Indeed, the ${2-}$body correlation function
$\mathcal{C}$ tends to its asymptotic value on a timescale of the order of the
dynamical time $t_{\rm dyn}$, while the ${1-}$body DF $F$ evolves on the secular
timescale ${ N t_{\rm dyn} }$ much larger. As a consequence, on the timescale of
evolution of $\lambda_{2}$, we may assume $F$ and $\lambda_{1}$ to be constant,
and at a given secular time $t$, $\lambda_{2}$ can be considered as equal to the
asymptotic value associated with the current frozen values of $\lambda_{1}$ and
$F$. In equation~\eqref{E_0_rewriting}, we may therefore assume that
$\lambda_{1}$ is frozen and that only $\lambda_{2}$ depends on time.

To simplify the calculations, we rely on the fact that the angles
$\bm{\theta}$ are ${ 2\pi-}$periodic. We define the Fourier transform w.r.t. the
angles $\bm{\theta}$ as
\begin{equation}
f (\bm{\theta} , \bm{J}) = \! \sum_{\bm{m}}  f_{\bm{m}} (\bm{J}) \, \re^{\ri \bm{m} \cdot \bm{\theta}} \;\;\; ; \;\;\; f_{\bm{m}} (\bm{J}) = \frac{1}{(2 \pi)^{d}} \!\! \int \!\! \rd \bm{\theta} \, f (\bm{\theta} , \bm{J}) \, \re^{- \ri \bm{m} \cdot \bm{\theta}} \, ,
\label{definition_Fourier_angles}
\end{equation}
where $d$ is the dimension of the considered physical space (e.g. ${d \!=\! 2}$
for a razor-thin disc as in~\citep{FouvryPichonChavanis2015,FouvryPichonMagorrianChavanis2015}).
Following~\citep{LyndenBell1994,Pichon1994,Chavanis2013}, we may Fourier
transform the interaction potential $U$ as
\begin{equation}
U (\bm{x} (\bm{\theta}_{1} , \bm{J}_{1}) \!-\! \bm{x} (\bm{\theta}_{2} , \bm{J}_{2})) = \sum_{\bm{m}_{1} , \bm{m}_{2}} A_{\bm{m}_{1} , \bm{m}_{2}} (\bm{J}_{1} , \bm{J}_{2}) \, \re^{\ri (\bm{m}_{1} \cdot \bm{\theta}_{1} - \bm{m}_{2} \cdot \bm{\theta}_{2} )} \, ,
\label{development_U}
\end{equation}
where ${ \bm{m}_{1} , \bm{m}_{2} \!\in\! \mathbb{Z}^{d} }$ are integer vectors. The coefficients $A_{\bm{m}_{1} , \bm{m}_{2}}$ satisfy the symmetry relations
\begin{equation}
A_{\bm{m}_{2} , \bm{m}_{1}} (\bm{J}_{2} , \bm{J}_{1}) = A_{- \bm{m}_{1} , - \bm{m}_{2}} (\bm{J}_{1} , \bm{J}_{2}) = \big[ A_{\bm{m}_{1} , \bm{m}_{2}} (\bm{J}_{1} , \bm{J}_{2}) \big]^{*} \, .
\label{symmetry_A}
\end{equation}
Similarly, we also introduce the Fourier transform of $\lambda_{2}$ as
\begin{equation}
\lambda_{2} (\bm{\theta}_{1} , \bm{J}_{1} , \bm{\theta}_{2} , \bm{J}_{2}) = \sum_{\bm{m}_{1} , \bm{m}_{2}} \lambda_{\bm{m}_{1} , \bm{m}_{2}} (\bm{J}_{1} , \bm{J}_{2}) \, \re^{\ri (\bm{m}_{1} \cdot \bm{\theta}_{1} + \bm{m}_{2} \cdot \bm{\theta}_{2})} \, .
\label{lambda2_FT}
\end{equation}
Since $\lambda_{2}$ is real, it satisfies the symmetry property
\begin{equation}
\lambda_{- \bm{m}_{1} , - \bm{m}_{2}} (\bm{J}_{1} , \bm{J}_{2}) \!=\! \lambda_{\bm{m}_{1} , \bm{m}_{2}}^{*} (\bm{J}_{1} , \bm{J}_{2}) \, .
\label{symmetry_lambda2}
\end{equation}
We now multiply equation~\eqref{E_0_rewriting} by ${ 1/(2 \pi)^{2 d} \re^{\ri (\bm{m}_{1} \cdot \bm{\theta}_{1} - \bm{m}_{2} \cdot \bm{\theta}_{2})} }$ and integrate it w.r.t. $\bm{\theta}_{1}$ and $\bm{\theta}_{2}$. The constraint ${ E (\Gamma_{1} , \Gamma_{2}) \!=\! 0 }$ then takes the form
\begin{equation}
\frac{\partial \lambda_{-\bm{m}_{1} , \bm{m}_{2}}}{\partial t} - \ri \Delta \omega \lambda_{ - \bm{m}_{1} , \bm{m}_{2}} = - \ri \mu A_{\bm{m}_{1} , \bm{m}_{2}}^{*} \bigg[ \bm{m}_{1} \!\cdot\! \frac{\partial \lambda_{1}}{\partial \bm{J}_{1}} \!-\! \bm{m}_{2} \!\cdot\! \frac{\partial \lambda_{1}}{\partial \bm{J}_{2}} \bigg] \, .
\label{E_TF_angles}
\end{equation}
where we used the shortening notations ${ \lambda_{ - \bm{m}_{1} , \bm{m}_{2}} \!=\! \lambda_{- \bm{m}_{1} , \bm{m}_{2}} (\bm{J}_{1} , \bm{J}_{2}) }$, ${ A_{\bm{m}_{1} , \bm{m}_{2}} \!=\! A_{\bm{m}_{1} , \bm{m}_{2}} (\bm{J}_{1} , \bm{J}_{2}) }$, and ${ \Delta \omega \!=\! \bm{m}_{1} \!\cdot\! \bm{\Omega}_{1} \!-\! \bm{m}_{2} \!\cdot\! \bm{\Omega}_{2} }$. Thanks to the adiabatic approximation, $\lambda_{1}$ can be assumed to be frozen, so that the differential equation~\eqref{E_TF_angles} can be straightforwardly solved. We recall that to obtain equation~\eqref{definition_E}, we had imposed the boundary condition ${ \lambda_{2} (T) \!=\! 0 }$, so that equation~\eqref{E_TF_angles} leads to
\begin{equation}
\lambda_{- \bm{m}_{1} , \bm{m}_{2}} (t) = \mu A_{\bm{m}_{1} , \bm{m}_{2}}^{*} \bigg[ \bm{m}_{1} \!\cdot\! \frac{\partial \lambda_{1}}{\partial \bm{J}_{1}} \!-\! \bm{m}_{2} \!\cdot\! \frac{\partial \lambda_{2}}{\partial \bm{J}_{2}} \bigg] \, \frac{1 \!-\! \re^{\ri \Delta \omega (t - T)}}{\Delta \omega} \, .
\label{solution_temporal_lambda2}
\end{equation}
At this stage, we assume that the arbitrary temporal bound $T$ is large compared to the considered time $t$, so as to consider only the forced regime of evolution, uninfluenced by the temporal boundary condition on $\lambda_{2}$. We may therefore place ourselves in the limit ${ T \!\to\! + \infty }$. Recalling the formula
\begin{equation}
\lim\limits_{T \to + \infty} \frac{\re^{\ri T \Delta \omega} - 1}{\Delta \omega} = \ri \pi \delta_{\rm D} (\Delta \omega) \, ,
\label{formula_asymptotics}
\end{equation}
equation~\eqref{solution_temporal_lambda2} immediately gives
\begin{align}
\lim\limits_{T \to + \infty} \lambda_{- \bm{m}_{1} , \bm{m}_{2}} (t) = \ri \, \pi \, \mu \, A_{\bm{m}_{1} , \bm{m}_{2}}^{*} \bigg[ \bm{m}_{1} \!\cdot\! \frac{\partial \lambda_{1}}{\partial \bm{J}_{1}} \!-\! \bm{m}_{2} \!\cdot\! \frac{\partial \lambda_{1}}{\partial \bm{J}_{2}} \bigg] \, \delta_{\rm D} (\bm{m}_{1} \!\cdot\! \bm{\Omega}_{1} \!-\! \bm{m}_{2} \!\cdot\! \bm{\Omega}_{2})  \, .
\label{limit_T}
\end{align}
Thanks to Bogoliubov's Ansatz, we have therefore inverted the constraint ${ E[F, \lambda_{1} , \lambda_{2}] \!=\! 0 }$ from equation~\eqref{E_0_rewriting}, so as to obtain ${ \lambda_{2} \!=\! \lambda_{2} [F , \lambda_{1}] }$. This will allow us to recover the expression of the Landau collision operator, as detailed in the next section.

\subsection{Recovering the inhomogeneous Landau operator}
\label{sec:recoverLandau}

After having inverted equation~\eqref{E_0_rewriting},
the expression of ${ \lambda_{2} \!=\! \lambda_{2} [F , \lambda_{1}] }$ may be substituted back in
equation~\eqref{rewriting_functional_all}. In equation~\eqref{rewriting_functional_all}, it then only remains the quantities $F$ and $\lambda_{1}$, and we note as ${ S [ F , \lambda_{1} ] }$ the remaining action term. It takes the form
\begin{equation}
S [ F , \lambda_{1} ] = \ri \!\! \int \!\! \rd t \rd \Gamma_{1} \,
\lambda_{1} A_{1} F + \frac{\ri}{2} \!\! \int \!\! \rd t \rd
\Gamma_{1} \rd \Gamma_{2} \, \lambda_{2} [F , \lambda_{1}] \,  G (\Gamma_{1} , \Gamma_{2}) \, .
\label{definition_Slast}
\end{equation}
Starting from the expressions~\eqref{A1_action} and~\eqref{G_action} of $A_{1}$ and $G$, and using the Fourier transform in angles introduced in equation~\eqref{definition_Fourier_angles}, we may rewrite the action from equation~\eqref{definition_Slast} as
\begin{equation}
S [ F , \lambda_{1} ] = \ri \!\! \int \!\! \rd t \rd \Gamma_{1} \, \lambda_{1} \frac{\partial F}{\partial t} - \frac{\ri}{2} \!\! \int \!\! \rd t \rd \Gamma_{1} \rd \Gamma_{2} \, \!\! \sum_{\bm{m}_{1} , \bm{m}_{2}} \bigg[ \ri \bm{m}_{1} \!\cdot\! \frac{\partial F}{\partial \bm{J}_{1}} F (\bm{J}_{2}) \, A_{\bm{m}_{1} , \bm{m}_{2}} \lambda_{- \bm{m}_{1} , \bm{m}_{2}} \!+\! \ri \bm{m}_{2} \!\cdot\! \frac{\partial F}{\partial \bm{J}_{2}} F (\bm{J}_{1}) \, \big[ A_{\bm{m}_{1} , \bm{m}_{2}} \lambda_{- \bm{m}_{1} , \bm{m}_{2}} \big]^{*} \bigg] \, ,
\label{rewrite_Slast}
\end{equation}
where, for simplicity, we used the notation ${ A_{\bm{m}_{1} , \bm{m}_{2}} \!=\!
A_{\bm{m}_{1} , \bm{m}_{2}} (\bm{J}_{1} , \bm{J}_{2}) }$, and ${ \lambda_{-
\bm{m}_{1} , \bm{m}_{2}} \!=\! \lambda_{- \bm{m}_{1} , \bm{m}_{2}} (\bm{J}_{1} ,
\bm{J}_{2}) }$. Using the symmetry properties from equations~\eqref{symmetry_A}
and~\eqref{symmetry_lambda2}, equation~\eqref{rewrite_Slast} immediately becomes
\begin{equation}
S [ F , \lambda_{1} ] = \ri \!\! \int \!\! \rd t \rd \Gamma_{1} \, \lambda_{1} \frac{\partial F}{\partial t} + \frac{\ri}{2} \!\! \int \!\! \rd t \rd \Gamma_{1} \rd \Gamma_{2} \, \!\! \sum_{\bm{m}_{1} , \bm{m}_{2}} \! \text{Im} \bigg[ A_{\bm{m}_{1} , \bm{m}_{2}} \lambda_{- \bm{m}_{1} , \bm{m}_{2}} \bigg] \, \bigg[ \bm{m}_{1} \!\cdot\! \frac{\partial F}{\partial \bm{J}_{1}} F (\bm{J}_{2}) - \bm{m}_{2} \!\cdot\! \frac{\partial F}{\partial \bm{J}_{2}} F (\bm{J}_{1}) \bigg] \, .
\label{rewrite_Slast_II}
\end{equation}
Thanks to the inversion from equation~\eqref{limit_T}, one can write
\begin{equation}
\text{Im} \bigg[ A_{\bm{m}_{1} , \bm{m}_{2}} \lambda_{ - \bm{m}_{1} , \bm{m}_{2}} \bigg] = \pi \, \mu \, \delta_{\rm D} (\bm{m}_{1} \!\cdot\! \bm{\Omega}_{1} \!-\! \bm{m}_{2} \!\cdot\! \bm{\Omega}_{2}) \, \big| A_{\bm{m}_{1} , \bm{m}_{2}} \big|^{2} \, \bigg[ \bm{m}_{1} \!\cdot\! \frac{\partial \lambda_{1}}{\partial \bm{J}_{1}} \!-\! \bm{m}_{2} \!\cdot\! \frac{\partial \lambda_{1}}{\partial \bm{J}_{2}} \bigg] \, .
\label{lambda2_inverted_real_im}
\end{equation}
Inserting this result in equation~\eqref{rewrite_Slast_II}, we get
\begin{align}
S [F , \lambda_{1}] = & \, \ri \!\! \int \!\! \rd t \rd \Gamma_{1} \, \lambda_{1} \frac{\partial F}{\partial t} \nonumber
\\
& \, \!\!\!\!\!\!\!\!\!\! \!+\! \frac{\ri}{2} \!\! \int \!\! \rd t \rd \Gamma_{1} \rd \Gamma_{2} \, \!\!\! \sum_{\bm{m}_{1} , \bm{m}_{2}} \!\! \pi \mu \, \delta_{\rm D} (\bm{m}_{1} \!\cdot\! \bm{\Omega}_{1} \!-\! \bm{m}_{2} \!\cdot\! \bm{\Omega}_{2}) \big| A_{\bm{m}_{1} , \bm{m}_{2}} \big|^{2} \bigg[ \bm{m}_{1} \!\cdot\! \frac{\partial \lambda_{1}}{\partial \bm{J}_{1}} \!-\! \bm{m}_{2} \!\cdot\! \frac{\partial \lambda_{1}}{\partial \bm{J}_{2}} \bigg] \bigg[ \bm{m}_{1} \!\cdot\! \frac{\partial F}{\partial \bm{J}_{1}} F (\bm{J}_{2}) \!-\! \bm{m}_{2} \!\cdot\! \frac{\partial F}{\partial \bm{J}_{2}} F (\bm{J}_{1}) \bigg] \, .
\label{injection_Slast}
\end{align}
The last step of the calculation is then to rewrite the second term in equation~\eqref{injection_Slast} under the form ${ \!\! \int \!\! \rd t \rd \Gamma_{1} \lambda_{1} (\Gamma_{1}) \, ... }$. This is straightforward thanks to an integration by parts. In the second term of equation~\eqref{injection_Slast}, let us focus on the term associated with ${ \bm{m}_{1} \!\cdot\! \partial \lambda_{1} / \partial \bm{J}_{1} }$. It reads
\begin{align}
 & \, \frac{\ri}{2} \pi \mu \!\! \int \!\! \rd t \rd \Gamma_{1} \rd \Gamma_{2} \, \!\! \sum_{\bm{m}_{1} , \bm{m}_{2}} \!\! \big| A_{\bm{m}_{1} , \bm{m}_{2}} \big|^{2} \delta_{\rm D} (\bm{m}_{1} \!\cdot\! \bm{\Omega}_{1} \!-\! \bm{m}_{2} \!\cdot\! \bm{\Omega}_{2}) \, \bm{m}_{1} \!\cdot\! \frac{\partial \lambda_{1}}{\partial \bm{J}_{1}} \bigg[ \bm{m}_{1} \!\cdot\! \frac{\partial F}{\partial \bm{J}_{1}} F (\bm{J}_{2}) \!-\! \bm{m}_{2} \!\cdot\! \frac{\partial F}{\partial \bm{J}_{2}} F (\bm{J}_{1}) \bigg] \nonumber
\\
& \, = - \frac{\ri}{2} \pi (2 \pi)^{d} \mu \!\!\! \int \!\!\! \rd t \rd \Gamma_{1} \lambda_{1} (\bm{J}_{1}) \frac{\partial }{\partial \bm{J}_{1}} \!\cdot\! \bigg[ \!\! \sum_{\bm{m}_{1} , \bm{m}_{2}} \!\! \bm{m}_{1} \!\!\! \int \!\!\! \rd \bm{J}_{2} \, \delta_{\rm D} (\bm{m}_{1} \!\cdot\! \bm{\Omega}_{1} \!-\! \bm{m}_{2} \!\cdot\! \bm{\Omega}_{2}) \big| A_{\bm{m}_{1} , \bm{m}_{2}} \big|^{2} \, \bigg[ \bm{m}_{1} \!\cdot\! \frac{\partial F}{\partial \bm{J}_{1}} F (\bm{J}_{2}) \!-\! \bm{m}_{2} \!\cdot\! \frac{\partial F}{\partial \bm{J}_{2}} F (\bm{J}_{1}) \bigg] \bigg] \, ,
\label{integration_by_parts_lambda1}
\end{align}
where the additional prefactor ${ (2 \pi)^{d} }$ comes from the transformation ${ \!\! \int \!\! \rd \Gamma_{2} \, f (\bm{J}_{2}) \!=\! (2 \pi)^{d} \!\! \int \!\! \rd \bm{J}_{2} \, f (\bm{J}_{2}) }$. One can perform the exact same calculation for the second term present in equation~\eqref{injection_Slast} associated with ${ \bm{m}_{2} \!\cdot\! \partial \lambda_{1} / \partial \bm{J}_{2} }$. One only has to permute the indices ${ 1 \!\leftrightarrow\! 2 }$, and recovers the exact same contribution as in equation~\eqref{integration_by_parts_lambda1}. As a consequence, one can get rid of the factor ${ 1/2 }$ present in equation~\eqref{injection_Slast}, so that it becomes
\begin{align}
S [F , \lambda_{1}] & \, =  \ri \!\! \int \!\! \rd t \rd \Gamma_{1} \, \lambda_{1} (\Gamma_{1}) \nonumber
\\
& \, \times  \bigg\{ \frac{\partial F}{\partial t} \!-\! \pi (2 \pi)^{d} \mu \frac{\partial }{\partial \bm{J}_{1}} \!\cdot\! \bigg[ \!\! \sum_{\bm{m}_{1} , \bm{m}_{2}} \!\! \bm{m}_{1} \!\! \int \!\! \rd \bm{J}_{2} \, \delta_{\rm D} (\bm{m}_{1} \!\cdot\! \bm{\Omega}_{1} \!-\! \bm{m}_{2} \!\cdot\! \bm{\Omega}_{2}) \big| A_{\bm{m}_{1} , \bm{m}_{2}} \big|^{2} \, \bigg[ \bm{m}_{1} \!\cdot\! \frac{\partial F}{\partial \bm{J}_{1}} F (\bm{J}_{2}) \!-\! \bm{m}_{2} \!\cdot\! \frac{\partial F}{\partial \bm{J}_{2}} F(\bm{J}_{1}) \bigg] \bigg] \bigg\} \, .
\label{final_Slast}
\end{align}
By integrating functionally equation~\eqref{final_Slast} w.r.t. $\lambda_{1}$, one finally obtains the expression of the inhomogeneous Landau equation which reads
\begin{equation}
\frac{\partial F}{\partial t} = \pi (2 \pi)^{d} \, \mu \,  \frac{\partial }{\partial \bm{J}_{1}} \!\cdot\!  \bigg[ \!\sum_{\bm{m}_{1} , \bm{m}_{2}} \!\bm{m}_{1} \!\! \int \!\! \rd \bm{J}_{2} \, \delta_{\rm D} (\bm{m}_{1} \!\cdot\! \bm{\Omega}_{1} \!-\! \bm{m}_{2} \!\cdot\! \bm{\Omega}_{2}) \, \big| A_{\bm{m}_{1} , \bm{m}_{2}} (\bm{J}_{1} , \bm{J}_{2}) \big|^{2} \bigg( \bm{m}_{1} \!\cdot\! \frac{\partial }{\partial \bm{J}_{1}} \!-\! \bm{m}_{2} \!\cdot\! \frac{\partial }{\partial \bm{J}_{2}} \bigg) \, F (\bm{J}_{1} , t) \, F (\bm{J}_{2} , t) \bigg] \, .
\label{Landau_equation}
\end{equation}
As a conclusion using the functional integral approach presented in~\citep{JolicoeurGuillou1989}, we have been able to recover in an alternative manner the inhomogeneous Landau equation obtained in~\citep{Chavanis2013}.

\section{Conclusion}
\label{sec:conclusion}

Understanding the long-term evolution of astrophysical dynamical systems is a
subject of renewed interest. On galactic scales we now have the well established
$\Lambda$CDM paradigm for the formation of structures. It allows us to quantify
in detail the statistical impacts of cosmic perturbations on self-gravitating
systems. 
These developments offer new clues to address the pressing question of the respective long-term roles of nature vs. nurture in the establishment of the observed properties of these systems. Numerous dynamical challenges are therefore ready to be re-examined in much greater detail than before.
Examples include: the secular evolution of the metallicity-dispersion
relationship in galactic discs, the mechanisms of disc thickening by molecular
clouds and/or spiral waves, the stellar dynamical evolution of the Galactic
centre, the evolution of proto-planetary discs of debris, etc. Yet,
characterising the secular evolution of such  systems remains a difficult task
since it requires intricate inhomogeneous kinetic models, complex numerical
experiments, and an accurate physical understanding of the involved competing  physical
processes. Kinetic equations such as the Landau and Balescu-Lenard equations are
expected to provide a crucial new lightning on these complex dynamical
processes.

Using the functional integral formalism introduced in~\citep{JolicoeurGuillou1989}, we showed how one may derive through this approach the inhomogeneous Landau equation~\eqref{Landau_equation}.
This calculation offered new insights on the content of this kinetic equation.
A natural next step of this calculation would be to show how one may use the same method to derive the inhomogeneous Balescu-Lenard equation~\citep{Heyvaerts2010,Chavanis2012}. Such a derivation is expected to be more involved, because one has to take into account the polarisation dressing of the potential fluctuations. In the inhomogeneous context, this requires to rely on the matrix method~\citep{Kalnajs2} and to introduce potential-density elements. This will be the subject of a future work.
Finally, one could expect this new functional integral approach to be
applicable to other kinetic equations. For example, 
this methodology can be transposed to the kinetic theory of two-dimensional
point vortices~\cite{Chavanis2012Onsager}. On the other hand, because of its
alternative point of view, such a method may turn out fruitful to tackle the
question of obtaining a closed kinetic equation when higher order correlation
terms are taken into account.\footnote{This is important, in particular, for one
dimensional systems for which the Balescu-Lenard
collision term  vanishes in the homogeneous case, so that three-body
or higher correlation functions have to be considered.} This is also the topic
of ongoing work. More generally, it would be of great interest to identify in
which contexts this functional approach could be more successful.

\section*{Acknowledgements}
\label{sec:acknowledgements}

JBF, PHC, and CP thank the CNRS Inphyniti program for funding.
JBF and CP also thank the theoretical physics sub-department, Oxford, for hospitality and the CNRS-Oxford 
exchange program for funding.
We thank Francis Bernardeau, John Magorrian, and Simon Prunet for stimulating discussions.
This work is partially supported by the Spin(e) grants ANR-13-BS05-0005 of the French Agence Nationale de la Recherche
(\texttt{http://cosmicorigin.org}).

\bibliographystyle{elsarticle-num-names}
\bibliography{references.bib}

\end{document}